# The wetting problem of fluids on solid surfaces: Dynamics of lines and contact angle hysteresis

Henri Gouin

*Laboratoire de Modélisation en Mécanique et Thermodynamique, E.A. 2596,*
*Université d'Aix-Marseille, Case 322, Avenue Escadrille Normandie-Niemen,*
*13397 Marseille Cedex 20, France.*
Electronic-mail : henri.gouin@meca.u-3mrs.fr

**Abstract.** In 1805, Young was the first who introduced an expression for contact angle in static, but today, the motion of the contact-line formed at the intersection of two immiscible fluids and a solid is still subject to dispute. By means of the new physical concept of line viscosity, the equations of motions and boundary conditions for fluids in contact on a solid surface together with interface and contact-line are revisited. A new Young-Dupré equation for the dynamic contact angle is deduced. The interfacial energies between fluids and solid take into account the chemical heterogeneities and the solid surface roughness. A scaling analysis of the microscopic law associated with the Young-Dupré dynamic equation allows us to obtain a new macroscopic equation for the motion of the contact-line. Here we show that our theoretical predictions fit perfectly together with the contact angle hysteresis phenomenon and the experimentally well-known results expressing the dependence of the dynamic contact angle on the celerity of the contact-line. We additively get a quantitative explanation for the maximum speed of wetting (and dewetting).

## 1. INTRODUCTION

We propose a dynamic model of motion of the contact system between two fluids and a solid surface [1-3]. Comparisons between our results and recent experiments or behaviours from statistical physics show an excellent fit on the qualitative level as well as the quantitative one [4-6].
The equations of fluid motions are those of Navier-Stokes [7,8]. The boundary conditions are expressed at the same time on different surfaces by dynamic Laplace formulae, but also on the contact-line by a differential equation using the new term of line viscosity. This differential equation, namely the expression of the Young-Dupré condition of the dynamic contact angle, takes into account the line viscosity but also the inhomogeneousness of all types on the solid surface [9-12]. It gives the microscopic behaviour of the Young contact angle. The inhomogeneousness is distributed at distances going from a few tens to a hundred Angströms. This distance is that of the operating ranges of intermolecular forces which command the surface energies. With the lower part of this scale, energies are homogenized and distances are no longer significant on a macroscopic scale [13,14].
The hysteresis behaviour depends solely on the physico-chemical properties of the solid surface and on the celerity of the contact-line. Its universal form makes it possible to discuss the general case independently of any particular apparatus. All the efforts on the contact-line do not have the same effects. In our scaling, some of them produce a work that tends to zero and they do not appear in the macroscopic expression of the contact system motion. It is noteworthy that the weak differences between potentials have huge implications. This may seem surprising at first sight. It is due to the fact that the contact-line is massless and consequently its motion equation is not in the same form as for material systems [15].
We can note finally that the often proposed concepts of Young angles (apparent, measured, real etc.) [16,17] no longer appear in our model; only the expression of the angle associated with the macroscopic contact-line behaviour is meaningful in the framework of continuum mechanics. Jumps on the inhomogeneousness are also considered; a shift factor is proposed by Hoffman correcting the relation between the line celerity and the Young angle [7]. They appear unnecessary in our model. Lastly, other models using a notion of dynamic surface energy are also considered in the literature [18]. They are out from the framework of our study.

## 2. KINEMATICS OF A FLUID AT A CONTACT LINE AND DISSIPATIVE FUNCTION

To solve a boundary-value problem associated with the governing equations, it is necessary to detail what conditions should be imposed near a moving contact-line. The difficulty arises with the no-slip condition usually considered in viscous fluid problems. In fact two main problems crop up in the physics of contact-line.

The first one comes from the kinematics of the contact-line itself. It is thought there is a contradiction in assuming the no-slip condition at a solid wall and expecting the fluid to displace another one (for example a liquid displaces air). The classical concept of fluid kinematics does not satisfactorily describe the solid-fluid-fluid intersection region. In a larger fluid motion axiomatic, a no-slip boundary condition and a moving contact-line are kinematically compatible concepts. One fluid undergoes a rolling-sticking motion while the motion of the other fluid is more complex. No constitutive equation for the bulk fluids is imposed. The derived properties of the motion are characteristic of Newtonian as well as non-Newtonian fluids. The wall shape roughness is not excluded.

The key idea is that a fluid which does not slip on a solid surface does not preclude the possibility that at some instant a fluid material point may leave the surface: the velocity of the fluid must equal the solid velocity at the surface.

The representation is a consequence of experimental, theoretical and numerical data obtained in the literature. When the fluid advances, the particles of the fluid interface come and stick on the solid wall according to a movement of adhesion similar to an adhesive tape that one would stick on the wall. These various observations are schematized in a main case represented in the plane section in Fig. 1. The contact-line acts in a similar way as a "shock wave" and is not a material line: along the contact-line, the velocity of the fluid is discontinuous and we emphasize that the multi-valued velocity of the fluid at the contact-line comes from the fact that the two fluids stick a solid surface [4].

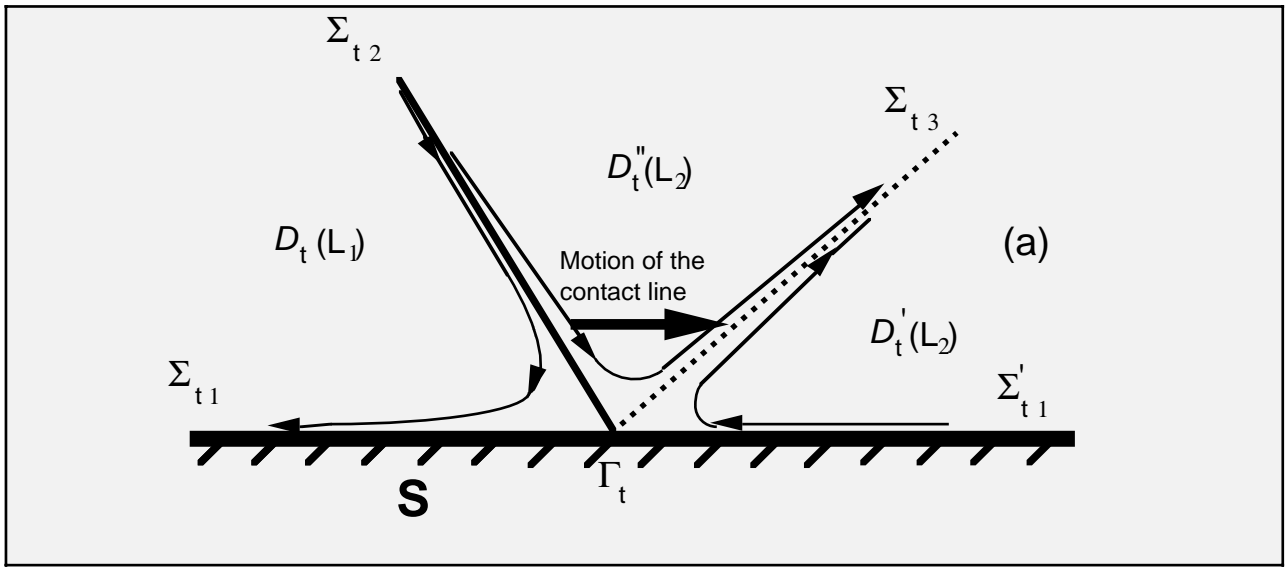

**Figure 1.** Typical motion of fluids in contact on a solid surface: the contact-line moves along the edge $\Sigma_{t1}$ of S towards $\Sigma'_{t1}$. The fluid $L_1$ is presented solely in one field $D_t(L_1)$ and moves along material surface $\Sigma_{t2}$ which has a common interface with the fluid $L_2$. The velocity of the fluid $L_1$ is equal to that of the fluid $L_2$ along the interface $\Sigma_{t2}$. The particles of the fluid $L_1$ belonging to $\Sigma_{t2}$ come into contact with S along the line $\Gamma_t$ and then adhere to $\Sigma_{t1}$. The fluid $L_2$ is made up of two independent fields $D'_t(L_2)$ and $D''_t(L_2)$. A common material surface $\Sigma_{t3}$ separates the two parts of the fluid $L_2$. The particles from $L_2$ belonging to $\Sigma_{t2}$ come into contact with S along $\Gamma_t$ then fork on $\Sigma_{t3}$ and move away from $\Gamma_t$. The particles of $L_2$ in contact with $\Sigma'_{t1}$ are reached by $\Gamma_t$ and subsequently move away from $\Gamma_t$ along $\Sigma_{t3}$. We obtain three other similar cases in exchanging fluids $L_1$ and $L_2$ and the direction of the motion.

The second problem arises from the dissipative function of a Newtonian fluid at the contact-line between two immiscible fluids and a solid boundary [19]. The condition of adherence generates a rate of energy dissipation which becomes infinite along the contact-line.

We will adopt the following framework:

At the distance e from the contact-line the viscosity stresses are those of Navier-Stokes and the liquid adheres to the wall. Near the contact-line (distance less e), the liquid slips on the wall and the flow is no longer Newtonian. On the solid surface S, the variations of the fluid velocities are along S and for a non isotropic case of a linear viscous stress tensor, the principal part of the viscous stress vector is in the form:

$$\mathbf{K}_V = -\nu \frac{\partial v^1}{\partial x} \, \mathbf{n}$$

where **n** is the unit normal vector at the contact-line along S, x the curvilinear coordinate orthogonal to the contact-line and $v^1$ the fluid velocity component along the x-coordinate.

Assuming that $v$ is a constant near the contact-line, we integrate $\mathbf{K}_V$ between point $P_e$ of curvilinear coordinate -e and point P on the contact-line. Then $\mathbf{T}_V = - \int_{-e}^{0} \mathbf{K}_V \, dx$ and the form of the line viscosity vector $\mathbf{T}_V$ for the linear model is

$$\mathbf{T}_V = \tau \, \mathbf{n} \quad \text{with} \quad \tau = -v \, u \tag{1}$$

where u is the contact-line celerity. This line viscosity vector $\mathbf{T}_V$ has the physical dimension of a force per unit of length.

Modern computers calculate the motion of a few thousands of molecules of liquid in the vicinity of the contact system. The slip condition is fully justified but only at a few Angströms distance e from the contact-line (one or two molecular distances). Beyond this zone, the liquid adheres to the wall and the Navier-Stokes equations seem to be a quite adapted model. In the e-zone of slipping, the tensor of viscosity is expressed by the single term $\mathbf{T}_V$ regrouping the non-Newtonian stresses of the contact system. Far from the critical conditions, the contact-line has a thickness of a few molecules and merges with the e-zone of slipping: on the scale of a continuous medium, the slip zone appears with a null thickness and merges into the contact-line.

The well-known paradox of the dissipation rate becoming infinite in the vicinity of the contact-line becomes a purely mathematical paradox. Indeed, the singularity of the dissipative function comes from a term in the form log(R/e) where R is a macroscopic length scale. This term becomes infinite when e tends to zero. Due to the fact that e is of an order of two molecular distances, log(R/e) remains bounded (for example with R/e = $10^{10}$, log(R/e) is lower than 30) and the singularity does not have a physical existence. For slow motions, the dissipative terms coming from the Newtonian dissipative rate and from the line viscosity $v$ are small enough for the flow to be considered as isothermal.

## 3. THE DYNAMIC YOUNG-DUPRE EQUATION

The Lagrange-d'Alembert principle of virtual works can be handled by the usual techniques of variation calculus. The equations of fluid motions are those of Navier-Stokes including the condition of adherence to the solid wall. The boundary conditions are expressed at the same time on different surfaces by dynamic Laplace formulae, but also on the contact-line by a new dynamic Young-Dupré equation using the new term $v$ of line viscosity:

$$\sigma_2 \cos \theta + \sigma_1 + v \, u = 0 \tag{2}$$

where $\sigma_1$ is the difference between the surface energies of the two fluids and the solid, $\sigma_2$ is positive and constant along the fluid-fluid interface, $\theta$ represents the Young angle associated with the fluid-fluid interface and the solid wall. To be in accordance with the second law of thermodynamics $v$ must be positive. It is immediately noticed that with the advance of the contact-line, u is positive and the dynamic angle is higher than the equilibrium angle. The result is reversed when u is negative.

## 4. A CLASSICAL EXAMPLE OF EXPERIMENT AND THE SCALING BEHAVIOUR OF THE CONTACT-LINE

The apparatus is a cylindar tube of radius "a" with a vertical symmetry axis [20]. A liquid fills the cylinder up to a position determined by a piston (Fig. 2a). The liquid is in contact with air. The air-liquid constant surface energy is $\sigma_2$. The wall is inhomogeneous with a surface energy $\sigma_1 = \sigma_{sl} - \sigma_{sv}$ (difference between the superficial energies of solid-liquid and solid-air) depending on the geometrical position of the meniscus. The value $\sigma_1$ is invariant around the vertical axis. The surface energy on the interval AP of the wall is such that:

$$\sigma_1 = \sigma_{10} - K \sin \left( \frac{z}{\varepsilon a} \right) \tag{3}$$

where $\sigma_{10}$ is a constant surface energy, $\varepsilon$ is a small dimensionless parameter and the scalar K is positive and smaller than $\sigma_{10}$. (This expression is only an explicit case. For any periodic smooth function, calculations are unaltered).

The methods of similarity make it possible to consider slow motions so that the dynamic Laplace equations have the simple form of the static case. Because of the symmetry of revolution, the meniscus may be considered as a spherical cap (such is the case for speeds lower than a few cm.s$^{-1}$ in capillary tube of diameter less than a few millimeters and for a liquid as the glycerol).

Some geometrical calculus yield the dynamic equation of the contact-line:

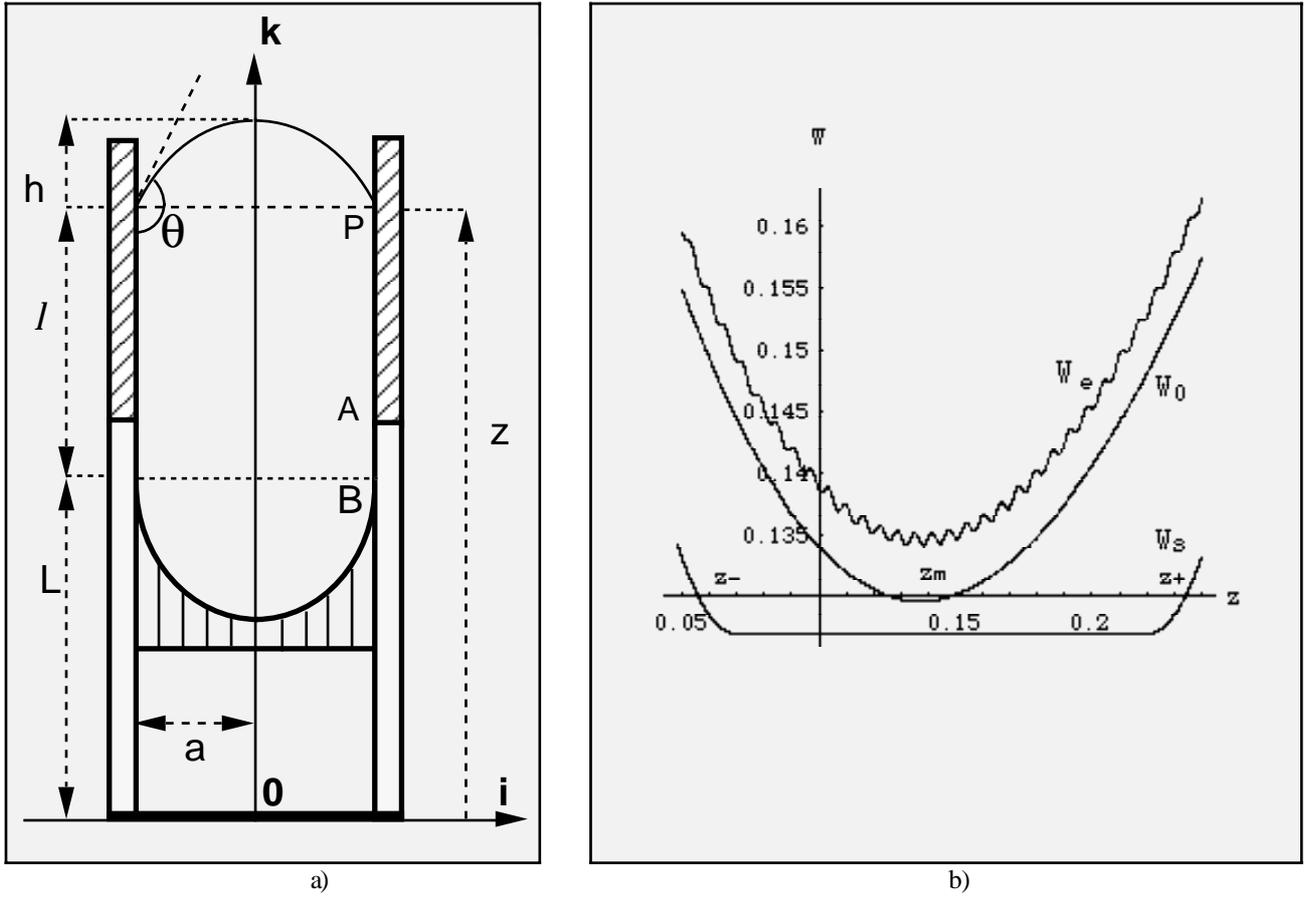

**Figures 2.** Figure a: "The apparatus". In the axisymmetric orthonormal axes O **i k**, z denotes the position of the contact-line, L(t) determines the position of the piston (z = $l$ + L); the Young angle between the meniscus and the wall is noted θ; h is the algebraic height of the meniscus.
Figure b: "Potentials associated with the motion equation of the contact-line". We trace the various potentials when the piston is at rest (z = $l$). In suitable units, we take the values a = 1, K = 0.005, $\sigma_{10}$ = − 0.5, $\sigma_2$ = 1. The curves are represented in the vicinity of the z-coordinate $z_m$ associated with $\theta_m$ = Arcos 0.5 (coordinate z is defined except for an additive constant). The potentials $W_o$, $W_e$ and $W_s$ have been shifted up or down (since potentials are defined except for an additive constant). Note that $W_o$ is the limit of $W_e$ when ε tends to 0, but it is not the same for $W_s$ which is basically different: tangent lines at the limiting points $z^-$ and $z^+$ are parallel to the z-axis and the function g(z) = − $\partial W_s/\partial z$ is of square root type at these points.

$$\nu \frac{dz}{dt} = F(z,t) + K \sin\left(\frac{z}{\varepsilon a}\right) \qquad (4)$$

$$\text{with } F(z,t) = \frac{2\lambda}{(1+\lambda^2)} \sigma_2 - \sigma_{10} \text{ and } \lambda = \frac{h}{a}.$$

We focus on the most characteristic case for which $\left|\frac{\sigma_{10}}{\sigma_2}\right| < 1$. We consider two significant cases encountered in experiments: piston at rest and piston in uniform motion (such as L(t) = $v_o$ t). Consequently,

$$\nu \frac{dl}{dt} = f(l) + K \sin\left(\frac{l + L(t)}{\varepsilon a}\right) - \nu v_o \qquad (5)$$

with $l$ = z − L(t) and f($l$) = F(z,t).
Differential equation (5) yields a single solution $l_\varepsilon$(t) when $l_\varepsilon$(0) = $l_o$ . We obtain the fundamental result that gives the behaviour of $l_\varepsilon$(t) when ε tends to zero [15].

### 4.1 Asymptotic behaviour of the contact-line

When ε tends to 0, $l_\varepsilon$(t) converges uniformly to the solution $l$(t) of the differential equation:

$$\nu \frac{dl}{dt} = g(l) \tag{6}$$

with $l(0) = l_o$ and $g(l) = \begin{cases} (f^2(l) - K^2)^{0.5} - \nu v_o & \text{if } l < l^- \\ -\nu v_o & \text{if } l^- < l < l^+ \\ -(f^2(l) - K^2)^{0.5} - \nu v_o & \text{if } l^+ < l \end{cases}$

where $l^-$ and $l^+$ are two constant abscissa verifying the relations $f(l^-) = K$ and $f(l^+) = -K$.
When the piston is fixed Eqs (5-6) become

$$\nu \frac{dl}{dt} = -\frac{dW_e}{dl} \tag{5'}$$

$$\nu \frac{dl}{dt} = -\frac{dW_s}{dl} \tag{6'}$$

with $W_e = W_o + W_1$ where $\frac{dW_o}{dl} = -f(l)$, $\frac{dW_1}{dl} = -K \sin(\frac{l}{\varepsilon a})$ and $\frac{dW_s}{dl} = -g(l)$.

Potentials $W_o$ and $W_s$ are convex functions.

### 4.1.1. The piston is fixed ($v_o = 0$).

The differential equation corresponding to the asymptotic behaviour of Eq. (5) when $\varepsilon$ tends to zero is associated with potential $W_s$. The driving force $g(l)$ has singularities of the square root type.

During the scaling - i.e. when $\varepsilon$ tends to zero - $W_s$ cannot be regarded as the limit of the sum of separate energies associated with the mean energy $W_o$ and the energy of perturbation ($W_e - W_o$). This shows that by changing the scale we lose the property of additivity of energy for the solution of the limit differential equation: $W_s$ is not the potential energy limit (Fig. 2b).

A physical interpretation of the previous limit behaviour can be given: when $\varepsilon$ tends to zero, the potential $W_e$ admits on the range $[l^-, l^+]$ a large number of local minima whose respective distances converge toward zero with $\varepsilon$. For any initial position $l_o$ in $[l^-, l^+]$ the closest local minimum is reached in an infinite time. When $\varepsilon$ tends to zero, any initial position $l_o$ of the interval $[l^-, l^+]$ is located between two local minima whose gap tends to zero with $\varepsilon$. On a macroscopic scale the local minima confound themselves with the initial value which becomes a "position of equilibrium".

Note that an angle $\theta_A(0) = \text{Arcos}(\frac{-\sigma_{10} - K}{\sigma_2})$ corresponds to the $l^-$-position. In the same way, an angle $\theta_R(0) = \text{Arcos}(\frac{-\sigma_{10} + K}{\sigma_2})$ corresponds to the $l^+$-position.

Then, $\theta_R(0) \leq \theta_m \leq \theta_A(0)$ with $\theta_m = \text{Arcos}(\frac{-\sigma_{10}}{\sigma_2})$.

This "static case" indicates that the Young angle $\theta$ is included in the interval $[\theta_R(0), \theta_A(0)]$. The final value of $\theta$ denoted $\theta_f$ depends on the initial position $l_o$ of $l$ and consequently on the initial value $\theta_o$ of $\theta$. We obtain the asymptotic behaviour.

$$\theta_o \in ]\theta_R(0), \theta_A(0)[ \implies \theta_f = \theta_o$$
$$\theta_o \leq \theta_R(0) \implies \theta_f = \theta_R(0)$$
$$\theta_o \geq \theta_A(0) \implies \theta_f = \theta_A(0)$$

### 4.1.2. Piston in uniform motion ($v_o \neq 0$).

It is easy to prove that for any initial condition $l_o$, a constant solution to Eq. (6) is reached in a finite time. We get $v_o = u$.
When the piston advances, $v_o = u > 0$ and when the piston recedes, $v_o = u < 0$. The constant solutions of Eq. (6) verify

$$f^2(l) - K^2 - \nu^2 u^2 = 0$$

The value of the advancing angle $\theta_A(u)$ is

$$\theta_A(u) = \text{Arcos} \left( \frac{-\sigma_{10} - (K^2 + \nu^2 u^2)^{0.5}}{\sigma_2} \right) \quad (7)$$

The value of the receding angle $\theta_R(u)$ is

$$\theta_R(u) = \text{Arcos} \left( \frac{-\sigma_{10} + (K^2 + \nu^2 u^2)^{0.5}}{\sigma_2} \right) \quad (8)$$

We deduce the inequalities

$$\theta_R(u) \leq \theta_R(0) \leq \theta_m \leq \theta_A(0) \leq \theta_A(u)$$

## 5. COMPARISON BETWEEN RESULTS, EXPERIMENTAL DATA AND OTHER MODELS OBTAINED IN THE LITERATURE

### 5.1 The dynamic line tension behaviour

Following the expression of the solid surface energy, the dynamic line tension admits a macroscopic behaviour associated with the differential equation (6). In the case of a piston at rest, the contact-line motion is governed by the equation

$$\nu \frac{dz}{dt} = \pm [(\sigma_2 \cos \theta + \sigma_{10})^2 - K^2]^{0.5} \quad (9)$$

where sign + or - depends on the direction of the line motion and $K = \max |\sigma_1 - \sigma_{10}|$ is the maximum of fluctuations in the fluid-solid energy with respect to its average value. The average value of the line tension due to surface energies is $\tau = \sigma_2 \cos \theta + \sigma_{10}$. The line tension $\tau$ has the dimension of a force by unit of length as $|\mathbf{T}_v|$ does. Then $\tau_m = -K$ represents the value of $\tau$ at $\theta = \theta_A(0)$.

If angle $\theta$ is close to $\theta_A(0)$, we obtain

$$\nu \frac{dz}{dt} \approx (2 \tau_m)^{0.5} (\tau - \tau_m)^{0.5} \quad (10)$$

If the contact angles are small while expanding $\theta$ to order 2, we obtain

$$\theta^2 - \theta_A^2(0) = \beta (\tau - \tau_m)$$

where $\beta$ is a suitable constant.

The results are extendable to the case of a piston in advancing motion. Eq. (10) is unchanged but the dynamic angle of contact is such that

$$(\sigma_2 \cos \theta + \sigma_{10})^2 - K^2 - \nu^2 u^2 = 0$$

If we note $\tau_u = -((\sigma_2 \cos \theta + \sigma_{10})^2 - \nu^2 u^2)^{0.5}$, the previous results are unchanged but $|\tau_u| < |\tau_m|$. In the case where the contact-line recedes it is easy to present calculations in the same way and to obtain analogous results.

These results agree with those obtained in the literature by molecular statistics or single defects [5,21].

### 5.2 Limit velocities of the contact-line

Limit velocities for advancing and receding contact-lines are revealed in experiments [22]. Indeed, the advancing dynamic Young angle must be lower than $\pi$. The associated celerity of the contact-line is $u_\pi$ and

$$\nu u_\pi = [(\sigma_2 - \sigma_{10})^2 - K^2]^{0.5} \quad (11)$$

In the same way the receding dynamic Young angle must be higher than 0. The associated celerity of the contact-line is $u_o$ and

$$\nu u_o = - [(\sigma_2 + \sigma_{10})^2 - K^2]^{0.5} \quad (12)$$

The knowledge of $\sigma_2$, $\sigma_{10}$, $u_o$ and $u_\pi$ allows us to determine K and $\nu$: the previous results make it possible to determine surface quality and line viscosity by simple measurements.

### 5.3 Connection between the dynamic contact angle, the line celerity and the line viscosity

The purpose is not to take the whole of the results obtained in the literature but to show by comparison with simple experiments that our model leads to qualitative and quantitative behaviour of the contact system.

The simplest way is to consider some surface inhomogeneousness given by a model represented with relation (3). The values of $\sigma_2$, $\sigma_{10}$ and K allow us to draw the graphs of applications given by (7) and (8). Only the ratio $\sigma_{10} / \sigma_2$ and the value of K are significant.

We present in Figures 3 experimental layouts drawn in the literature for an advancing motion of the contact-line. The similarity of the theoretical graph and experimental data is striking. It makes it possible to obtain the line viscosity value $\nu$ and the limit velocities $u_o$ and $u_\pi$.

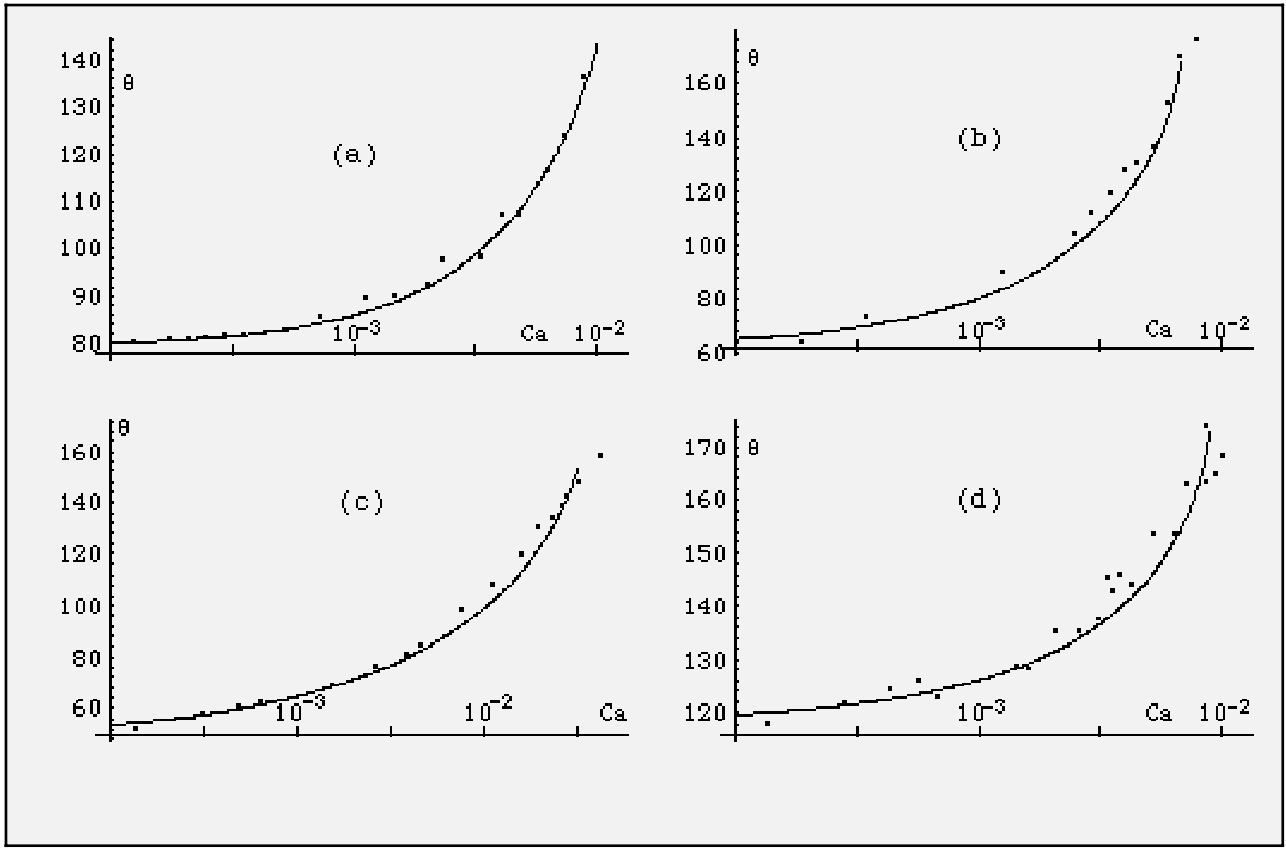

**Figures 3.** "Comparison between theoretical and experimental results". Due to the universality of our model, the macroscopic contact angle is plotted as a function of $Ca = \mu u / \sigma_2$. The Ca-axis is on a logarithmic scale. Plots from figures (a) and (b) are associated with the data of J.P. Stokes et al. (ref. [13] from [6]). Plots from figures (c) and (d) are associated with the data of G.M. Femigier and P. Jenffer (ref. [12] from [6]). Solid curves are ajusted with our results computed by Eq. (7). The solid surfaces are assumed to be smooth enough for the hysteresis to be small. It is easy to adjust the values of $\sigma_{10}$, $\sigma_2$ and $\nu$ to fit with experimental data denoted by points.

We use the experimental results for liquid flows in contact with air in capillary tubes. In the model suggested by relation (3), the solid surface inhomogeneousness is represented by the factor K. In fact, K $\ll \sigma_{10}$ and in relations (11) and (12) the term K can be neglected with regard to the terms $\sigma_2 - \sigma_{10}$ and $\sigma_2 + \sigma_{10}$. In order to be in experimental conditions, we suppose the diameter of the tube to be about 1 mm. Values of the viscosity and those of the liquid-air surface energy are given in textbooks. We deduce the numerical relations between the capillary number $Ca = \mu u / \sigma_2$ and u. These expressions are given in c.g.s. units. For water we obtain u = 7200 Ca and for glycerol u = 4.2 Ca.

The experimental curves yield limit velocities of the contact-line. They are associated in various measurements with the values of Ca ranging between $10^{-2}$ and $10^{-3}$. In the case of an air-glycerol interface, these values correspond to limit celerity $u_\pi$ included between 0.042 cm.s$^{-1}$ and 0.42 cm.s$^{-1}$ and in the case of an air-water interface to limit celerity $u_\pi$ included between 72 cm.s$^{-1}$ and 720 cm.s$^{-1}$.

For glycerol and a limit celerity of 0.42 cm.s$^{-1}$ and for a wetting static angle $\theta_A(0)$ of about 50° we obtain a line viscosity of about 240 c.g.s. In the case of water and for a limit celerity of 100 cm.s$^{-1}$, the line viscosity is about one c.g.s. unit. Due to the condition associated with the dimensionless Weber number, in order to obtain a spherical cap for the liquid-air interface, the case of water is obviously less realistic than the case of glycerol.

# 6. CONCLUSION

It is the first time a complete method of the wetting problem of the contact-line is presented in dynamics. The results are applicable in different fields as well as physics, physical-chemistry, biology and engineering.

The model proposed accounts for the phenomenon of hysteresis of the Young angle and expresses the movement of the line. Fluid motions are continuous along the contact system but velocities are discontinuous on the contact-line.

The microscopic motion of the contact system is governed by a simple differential equation. When the irregularities of liquid-wall surface energy vary over length that tends to zero relative to the size of the capillary tube, the solution of the microscopic motion tends to a limit which is the solution of a new differential equation. This differential equation of the macroscopic or homogenized motion is not the limit of the microscopic one.

The paradox of an infinite rate of dissipation in the contact system is solved: near the contact-line, the viscous stress tensor is replaced by a viscous tension of line.

The results of the example are independent of the tube radius. Indeed, the hysteresis behaviour depends solely on the physico-chemical properties of the solid surface and on the celerity of the contact-line. Its universal form makes it possible to discuss the general case independently of any particular apparatus. Relations (11) and (12) can be written without difficulty utilizing the inhomogeneousness of the solid surface in a different form from Eq. (3).

The most significant innovation is the introduction of the new concept of line viscosity. This term is essential to the construction of our model. Its value is obtained by using experimental measurements. It allows us to obtain the microscopic behaviour of the contact-line. The scaling of this behaviour to a macroscopic level yields the hysteresis phenomenon of the Young contact angle and the main experimental well-known results.

A complete study of our model in the wetting problem and the analysis of fluid motion near a contact-line are presented in [23].

## References

[1] Dussan V E.B., *Annual Rev. Fluid Mech.* **11** (1979) 371-400.
[2] de Gennes P.G., *Reviews of Modern Physics* **5** (1985) 827-863.
[3] Anderson D.M., McFadden G.B. & Wheeler A.A., *Annual Review Fluid Mech.* **30** (1997) 139-165.
[4] Dussan V E.B., Davis S.H., *J. Fluid Mech.* **65** (1974) 71-95.
[5] Koplick J., Banavar J.R. & Willemsen J.F., *Phys. Fluids* **A1** (1989) 781-794 .
[6] Zhou M.Y., Sheng P., *Phys. Rev. Letter* **64** (1990) 882-885.
[7] Hoffman R., *J. Colloid Interface Sci.* **50** (1975) 228-241.
[8] Cox R.G., *J. Fluid Mech.* **168** (1986) 169-200.
[9] Marmur A., *Colloid and Surfaces* **136** (1998) 81-88.
[10] Merchant G.J., Keller J.B., *Phys. Fluids* **A4** (1992) 477-485.
[11] Huh C., Mason S.G., *J. Colloid Interface Sci.* **60** (1977) 11-38.
[12] Wolansky G., Marmur A., *Colloid and Surfaces* **156** (1999) 381-388.
[13] Rowlinson J.S., Widom B., Molecular Theory of Capillarity. Clarendon Press, Oxford, 1984.
[14] Gouin H., *J. Phys. Chem.* **102** (1998) 1212-1218.
[15] Abeyaratne R., Chu C. & James R.D., *Philosophical Magazine A* **73** (1996) 457-497.
[16] Dussan V E.B., Ramé E. & Garoff S., *J. Fluid Mech.* **230** (1991) 97-116.
[17] Decker E.L., Frank B. , Sua Y. & Garoff S., *Colloid and Surfaces* **156** (1999) 177-189.
[18] Blake T.D., Bracke M. & Shikhmurzaev Y.D., *Physics of fluids* **11** (1999) 1995-2007.
[19] Pomeau Y., Pumir A., C. R. Acad. Sci. Paris, **299** (1984) 909-912.
[20] Dussan V E.B., *A.I.Ch.E.J.* **23** (1977) 131-132.
[21] Raphael E., de Gennes P.G., *J. Chem. Phys.* **90** (1989) 7577-7584.
[22] Blake T.D., Ruschak K.J., *Nature* **282** (1979) 489-491.
[23] Gouin H., *Continuum Mechanics and Thermodynamics*, (Submitted).